\documentclass[conference]{IEEEtran}
\IEEEoverridecommandlockouts
\usepackage{cite}
\usepackage{amsmath,amssymb,amsfonts}
\usepackage{algorithmic}
\usepackage[]{graphicx}
\usepackage{textcomp}
\usepackage{xcolor}
\usepackage{subcaption}
\usepackage{comment}
\def\BibTeX{{\rm B\kern-.05em{\sc i\kern-.025em b}\kern-.08em
    T\kern-.1667em\lower.7ex\hbox{E}\kern-.125emX}}
\begin{document}

\title{eSports Pro-Players Behavior \\During the Game Events: Statistical Analysis \\of Data Obtained Using the Smart Chair
}
\author{
	\IEEEauthorblockN{
		\textsuperscript{1,2}Anton Smerdov,
		\textsuperscript{1}Evgeny Burnaev,
		\textsuperscript{1,*}Andrey Somov
		}
	\IEEEauthorblockA{
	\textsuperscript{1}Skolkovo Institute of Science and Technology, CDISE, Moscow, Russia}
	\IEEEauthorblockA{
		\textsuperscript{2}Moscow Institute of Physics and Technology, Moscow, Russia}
	\IEEEauthorblockA{
		\textsuperscript{*}a.somov@skoltech.ru}
	}

\maketitle

\begin{abstract}
Today's competition between the professional eSports teams is so strong  that in-depth analysis of players' performance literally crucial for creating a powerful team. There are two main approaches to such an estimation: obtaining features and metrics directly from the in-game data or collecting detailed information about the player including data on his/her physical training. While the correlation between the  player's skill and in-game data has already been  covered in many papers, there are very few works related to analysis of eSports athlete's skill through  his/her physical behavior. We propose the smart chair platform which is to collect data on the person's behavior on the chair using an integrated accelerometer, a gyroscope and a magnetometer. We extract the important game events to define the players' physical reactions to them. The obtained data are used for training machine learning models in order to distinguish between the \textit{low-skilled} and  \textit{high-skilled} players. We extract and figure out the key features during the game and discuss the results.


\end{abstract}


\section{Introduction}


Nowadays eSports is a rapidly growing industry with more than billion players involved worldwide. The competition among top tier teams is strong and involves, apart from the teams themselves, their coaches, managers and associated scientists. As a result, eSports research has significantly progressed for the last few years.
But still, to the best of our knowledge there are few works related to the estimation of eSports athletes performance based on  their physical behaviour. The player reactions to the  game events can be investigated using Electroencephalography (EEG) \cite{eeg_cs} or brain waves \cite{dota_brain_waves}. Also, the game influence can be evaluated through the computer mouse movements \cite{affects_by_mouse}. 

Another approach for evaluating the athlete performance is based on the game statistics. Shim et al. research \cite{player_team_performance_study} is devoted to the calculation of Kill/Death/Assist (KDA), Kill Death Ratio (KDR) for a player in a first-person shooter game and predicting them for the next game rounds. First-person shooter research is reported in \cite{skill_capture_shooter} where the authors analyze  the dependence between the player skill and the data collected from the keyboard and mouse logs. On top of that there are works presenting the most important factors for players in Multiplayer Online Battle Arena (MOBA) \cite{skill_dota_2_1, skill_moba_2} and Massively Multiplayer Online Role-Playing Game (MMORPG) genres \cite{performance_motivation_mmorpg, mmorpg_mentoring_performance}.

Research into eSports include (i) affective computing~\cite{affective_comp}, (ii) prediction related research~\cite{prediction}, and (iii) social structures in teams~\cite{social_aspects}. Research in these areas suffers from purely theoretical research without proper experimental work with the professional eSports athletes. Indeed, data collection followed by modelling and interpretation tasks has the potential to forster the eSports research and make it practically feasible.



In this work, we use a \textit{smart chair} for data collection and further analysis of eSports athletes behaviour. Smart chairs have already been used in the  unobtrusive sensing applications: the pressure sensors embedded in a chair provide sufficient information about the sedentary patterns \cite{smart_chair_pressure_1, smart_chair_pressure_2}. This information can be used for making classification of the user activity \cite{chair_activity_1}, e.g. talking, coughing, eating, as well as stress detection \cite{chair_stress}. Another straightforward application is the user posture detection. Authors in \cite{smart_chair} report on the posture detection using the tilt sensors in addition to the pressure sensors. A similar method is used in \cite{the_most_useless_research} to access the user experience through the smart chair, and, in \cite{occupancy_detection} to build an occupancy detection system.

Another application of the smart chair concept is the  measurement of vital signs based on the heart related sensors. Authors in \cite{chair_bcg_2} and \cite{chair_bcg} used EMFi sensors integrated in the chair to measure the Ballistocardiogram (BCG). Ahn et al. \cite{chair_hrm} proposed the electrocardiography (ECG) method for the measurement of heart rate with  the sensors embedded in a chair.

A contribution of our paper is the collection of eSports data using the smart chair and further data analysis using machine learning algorithms. The important part of our work is extracting data about game events and using it to extract more meaningful features.

We managed to build the machine learning models which are able to classify the athletes on the basis of their skills. The best model demonstrates the 77\% accuracy and 0.88 ROC AUC score. The key features have been figured out and ranked.


This paper is organized as follows: in Section II we describe the smart chair platform used for data collection from the professional CS:GO athletes and amateur players. In Section III we present methodology used in this work. Data processing, feature extraction and machine learning algorithms are detailed in Section IV. We provide concluding remarks in Section V.

\section{Smart chair sensing platform}

\subsection{System architecture}

\begin{figure}[!bt]
	\centerline{\includegraphics[width=\linewidth]{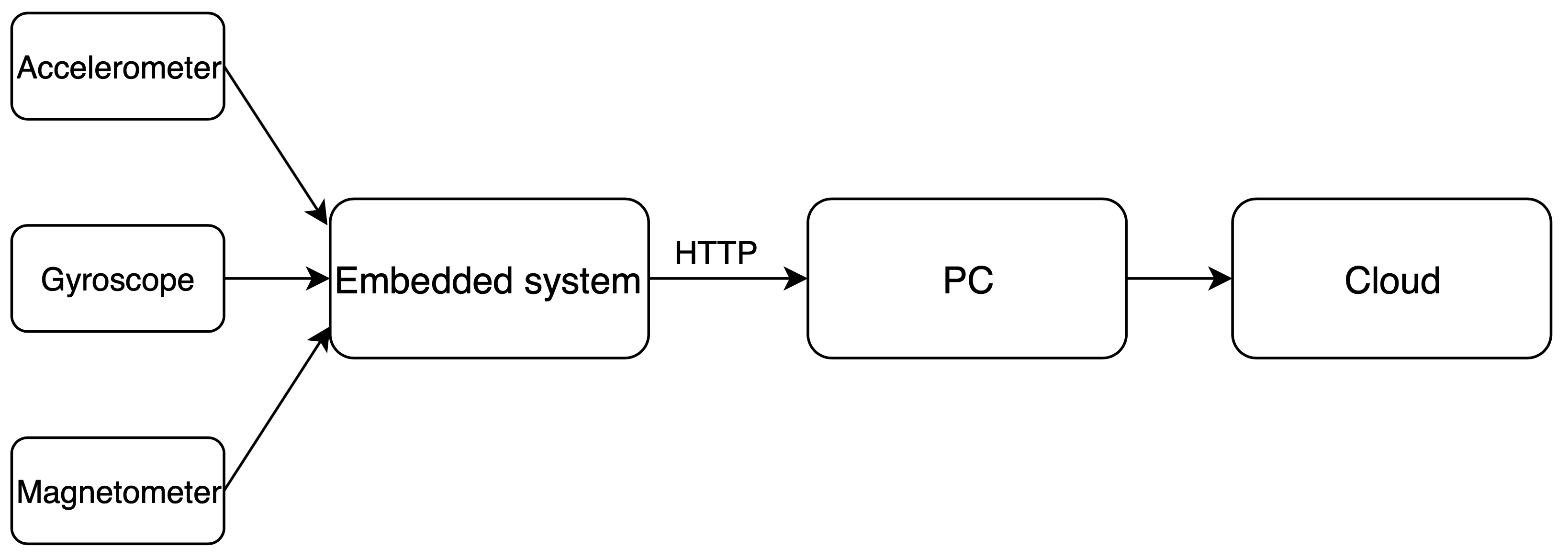}}
	\caption{System architecture.}
	\label{scheme}
\end{figure}

The smart chair platform consists of two units: a sensor unit for data collection and a server for data processing. The sensor unit consists of an  accelerometer, a magnetometer and a gyroscope. The data are collected by Motion Processing Unit (MPU) 9250 via I2C protocol.
 This unit is connected to a single-board computer Raspberry Pi 3. The platform can be further extended by adding some extra sensors, e.g. the pressure and temperature ones. The data are collected every 0.01 s. Upon collecting the data, the sensing unit makes a request and sends them to a server over the WiFi wireless channel. The data are sent to the server via the HTTP protocol every second in the JSON format. The data processing is realized on the server using the machine learning algorithms that are described in more details in the next section. The system architecture is shown in Fig.~\ref{scheme}.

\begin{figure}[!bt]
	\centering
	\begin{subfigure}[!tp]{0.5\textwidth}
	\centerline{\includegraphics[width=0.55\textwidth]{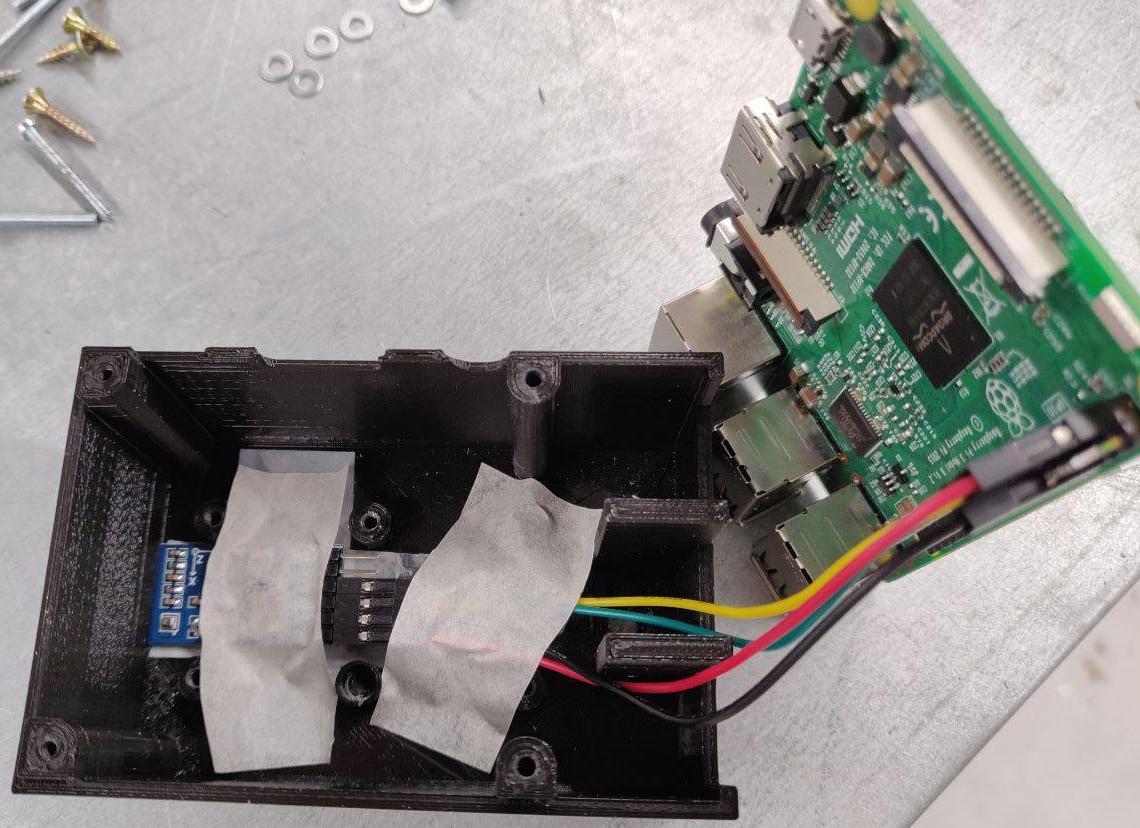}}
	\caption{Anatomy of the sensing system.}
	\label{sensors}
	\vspace{0.1cm}
	\end{subfigure}
	\newline
	\begin{subfigure}[!tp]{0.22\textwidth}
	\centerline{\includegraphics[height=4.8cm]{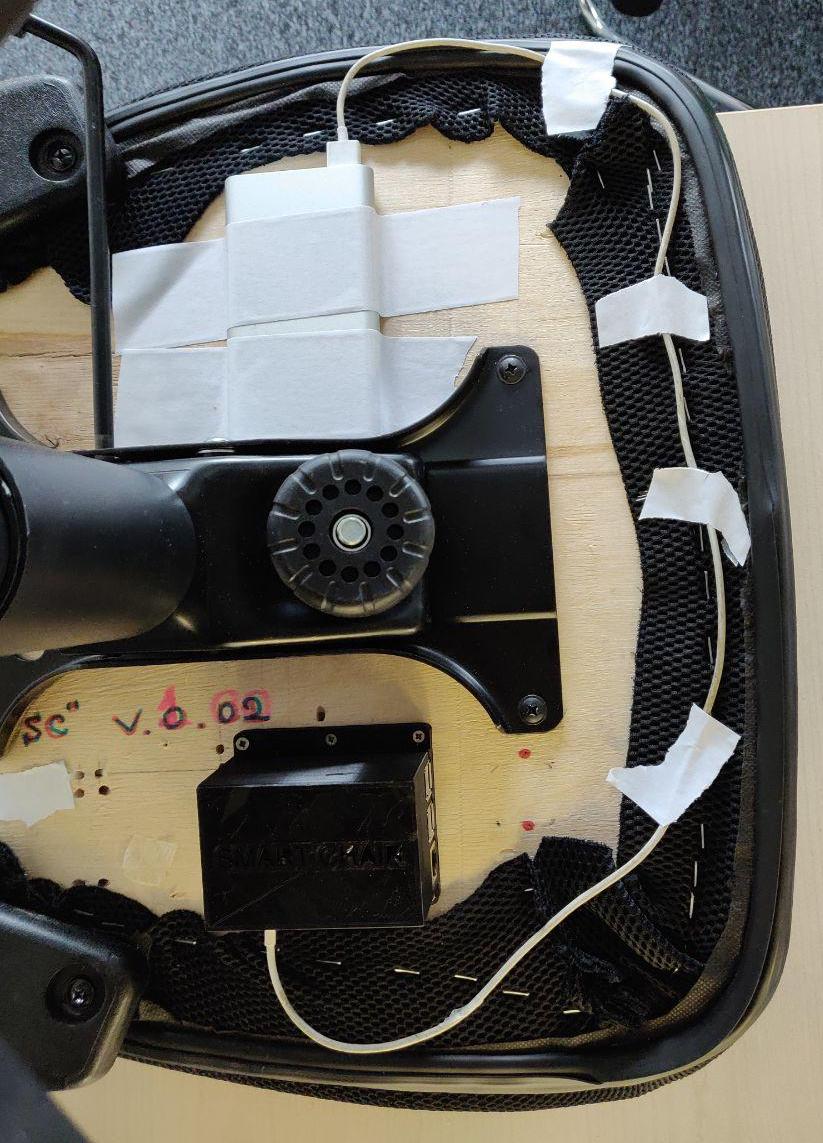}}
	\caption{Sensing module fixed on the bottom of a chair.}
	\label{chair}
	\end{subfigure}
	\begin{subfigure}[!tp]{0.22\textwidth}
	\centerline{\includegraphics[height=4.8cm]{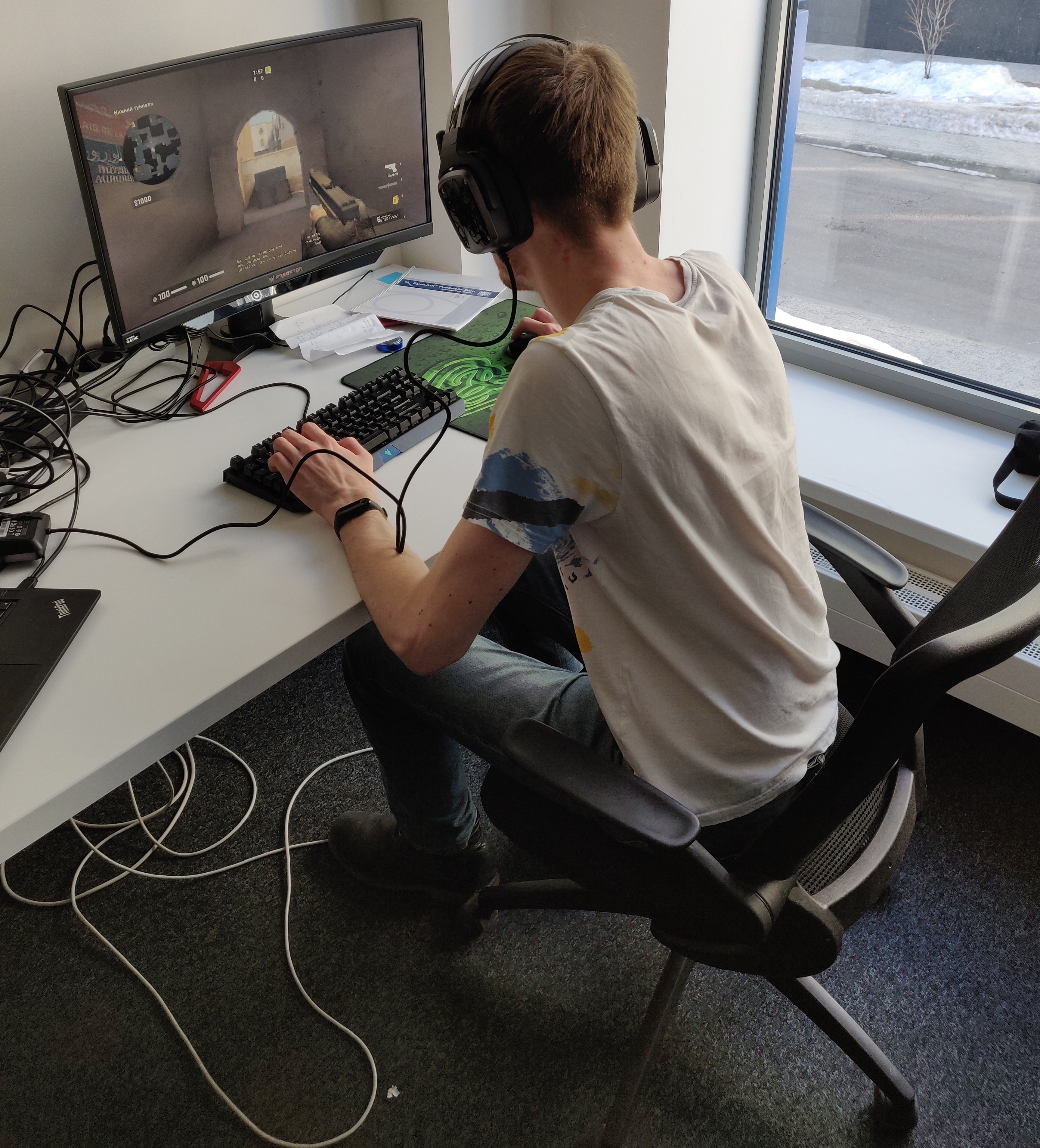}}
	\caption{Experimental testbed.\newline}
	\label{stand}
    \end{subfigure}
	\caption{Sensing system.}
	\label{sensing_system}
\end{figure}

The experimental testbed is shown in Fig~\ref{sensing_system}. The sensing unit is rigidly fixed at the bottom of the chair not to disturb the game process and capture all of the chair movements. In order to avoid redundant wires from the chair, we power it by a daily rechargeable external battery.


\begin{figure}[!bt]
	\centerline{\includegraphics[width=0.5\linewidth]{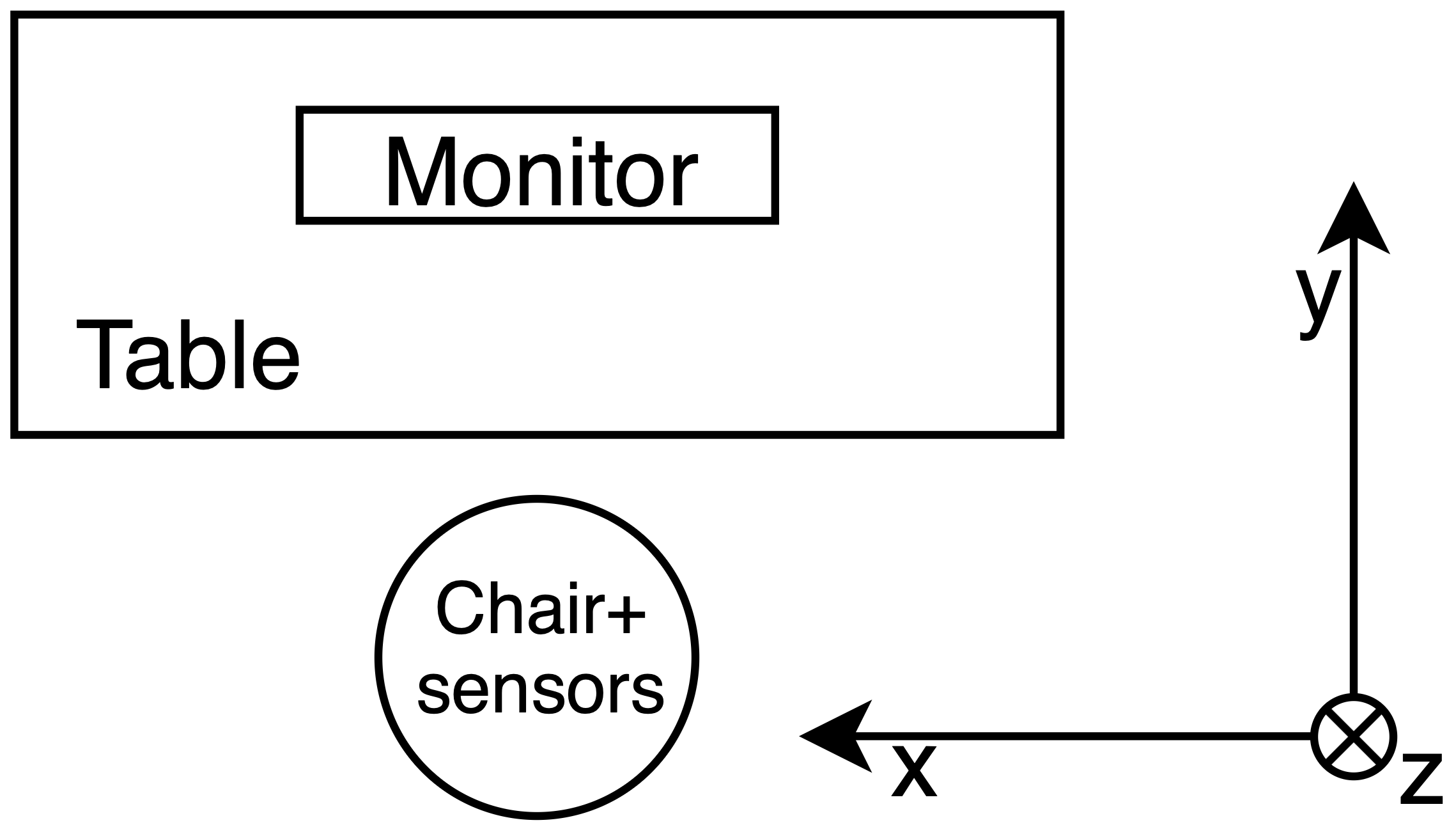}}
	\caption{Axes orientation.}
	\label{sensors_orientation}
\end{figure}

It is important to properly orient sensors properly to simplify further data processing. The axes orientation is shown in Fig.~\ref{sensors_orientation}. The axis \textit{z} is the vertical axis, $y$ is the axis passing through the player and the monitor, $x$ is the axis parallel to the gaming table.

\section{Methodology}

{

We have invited 9 professional athletes, primarily from the Monolith professional team specializing in CS:GO discipline, and 10 amateur players to take part in the experiment. Before the experiment we informed  all the participants about the project and the experiment procedure and collected their written consents. We asked the participants to fill in the questionnaire to make sure that they are in good form and do not take any drugs that might affect the experimental results.

In this work we apply the Retake modification of CS:GO discipline. In the Retake scenario a terrorist team plays against a counter-terrorist team. The terrorist team is made up of 2 players who typically play in a defensive manner: they have a bomb planted on the territory and have to defend it from the opposite team. The counter-terrorist team (3 players) is to to deactivate the bomb or, alternatively, kill their 'enemies' (the opposite team). The game user interface shows the bomb location on the map in the beginning of each round which lasts for approximately 40 s (there are 12 rounds altogether). The players have to buy the same set of weapons for each round. The Retake scenario is to be played continuously without any breaks between the rounds.
}

\section{Data Analysis}

\subsection{Data collection and Pre-processing}

{
	\it

}

We collected data from 19 persons playing CS:GO. Each gaming session lasts about 35 minutes. As a matter of fact, the player's behavior is sufficiently  characterized by a smaller timeseries.
That is why, in order to increase the amount of training data, we divide each player's log into 3-minutes sessions, up to 10 non-intersecting sessions per player. As a result, we got 171 labeled timeseries, some of them corresponding to the same players.

We also extracted the key game events from the gamelogs. For each session, moments of player killing and death were defined.
In order to capture the player's behaviour in the tough situations, the 'shootouts' events were extracted. These are the events when the player shoots  at least 3 times with the less than 3 seconds delay between the shots.

\begin{figure}[!bt]
	\centering
	\begin{subfigure}{0.45\textwidth}
	\centerline{\includegraphics[width=\linewidth]{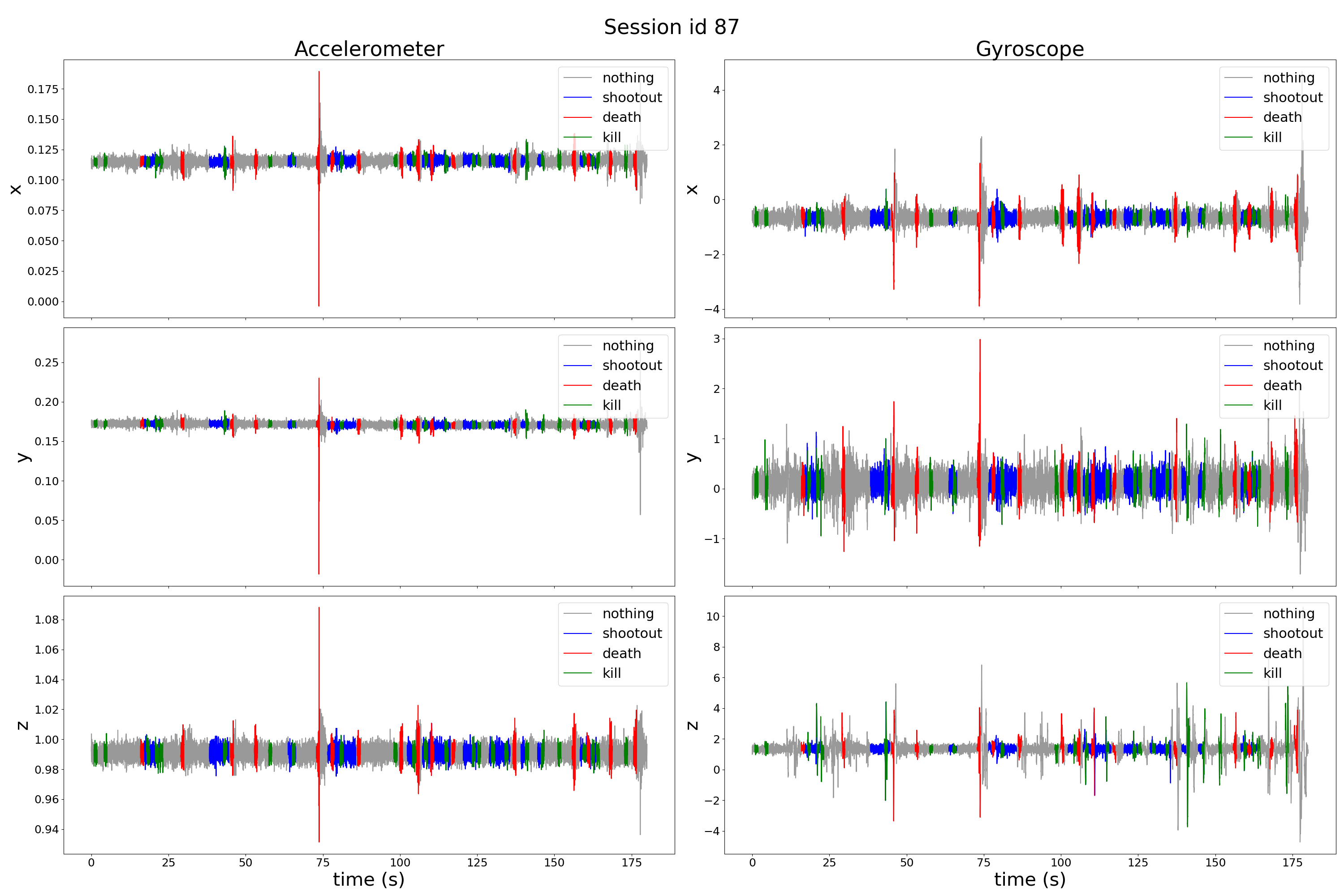}}
	\label{sensors_raw_1}
	\end{subfigure}
	\newline
	\begin{subfigure}{0.45\textwidth}
	\centerline{\includegraphics[width=\linewidth]{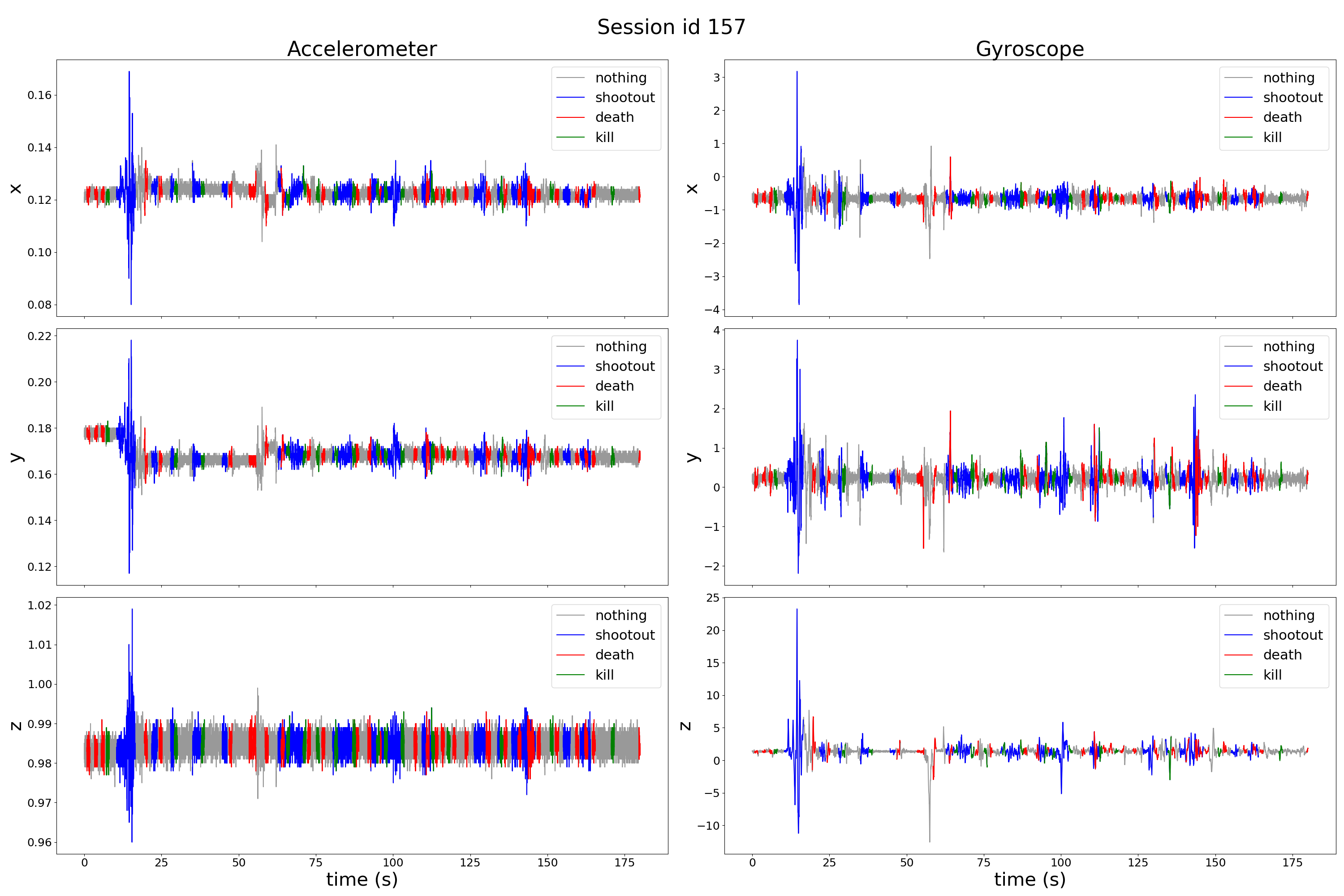}}
	\label{sensors_raw_2}
	\end{subfigure}
	\caption{Examples of the raw data collected from the accelerometer and gyroscope. Colors correspond to game events.}
	\label{sensors_raw}
\end{figure}

In our experiment we have three types of data: acceleration from the accelerometer, spatial orientation from the magnetometer and angular velocity from the gyroscope.
Examples of the raw data from the accelerometer and gyroscope for a 3-minute session are shown in Fig.~\ref{sensors_raw}. In order to demonstrate the correlation with the game events we colored the  corresponding moments.

It is clear that for most of the time the player does not make lot of movements while sitting on the chair. Some disturbance, however, constantly occurs, sometimes simultaneously with the key game events. Our goal is to check whether the reaction to the game events can describe the player's skill.

\begin{figure}[!b]
	\centering
	\begin{subfigure}{0.45\textwidth}
	\centerline{\includegraphics[width=\linewidth]{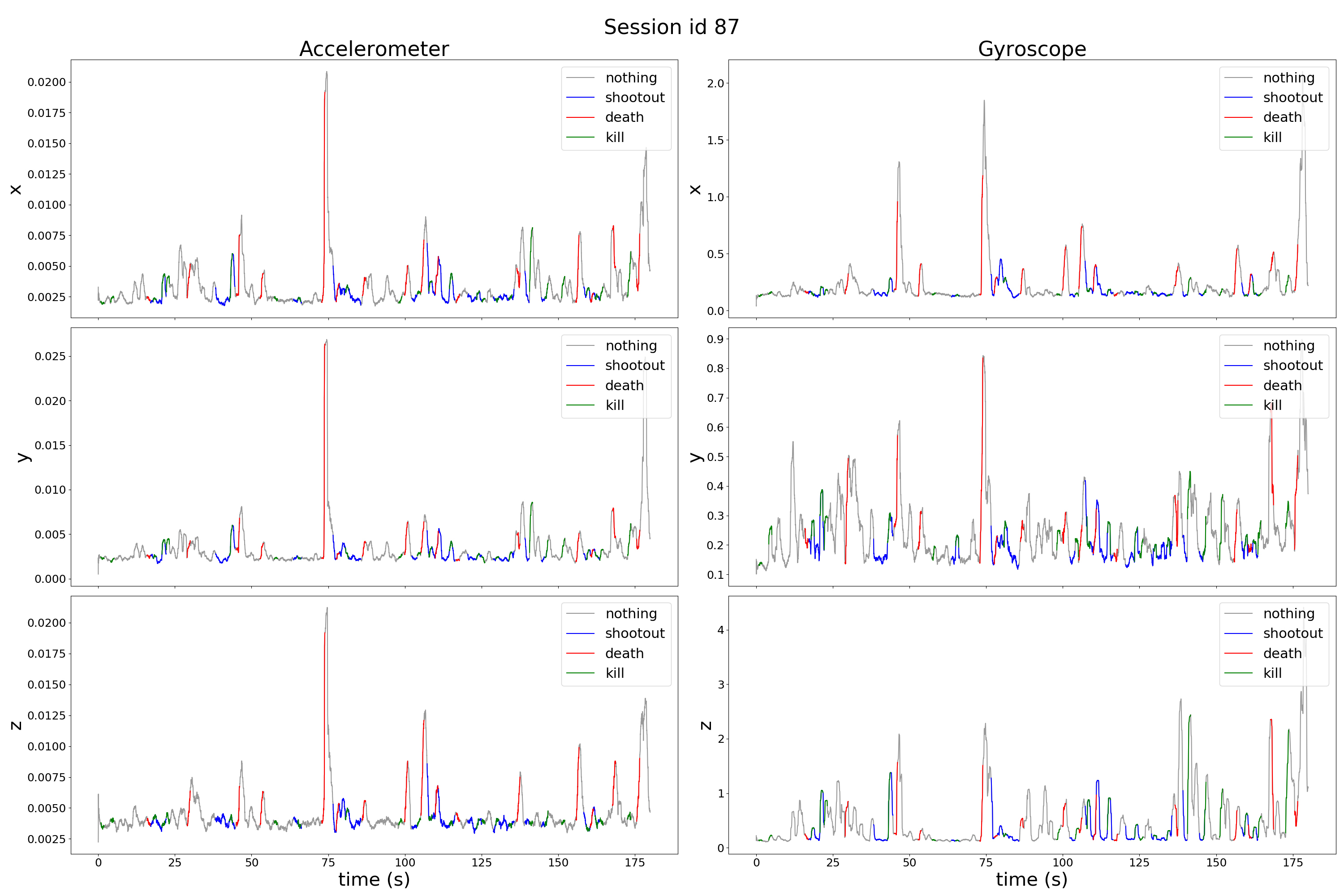}}
	\label{sensors_std_1}
	\end{subfigure}
	\newline
	\begin{subfigure}{0.45\textwidth}
	\centerline{\includegraphics[width=\linewidth]{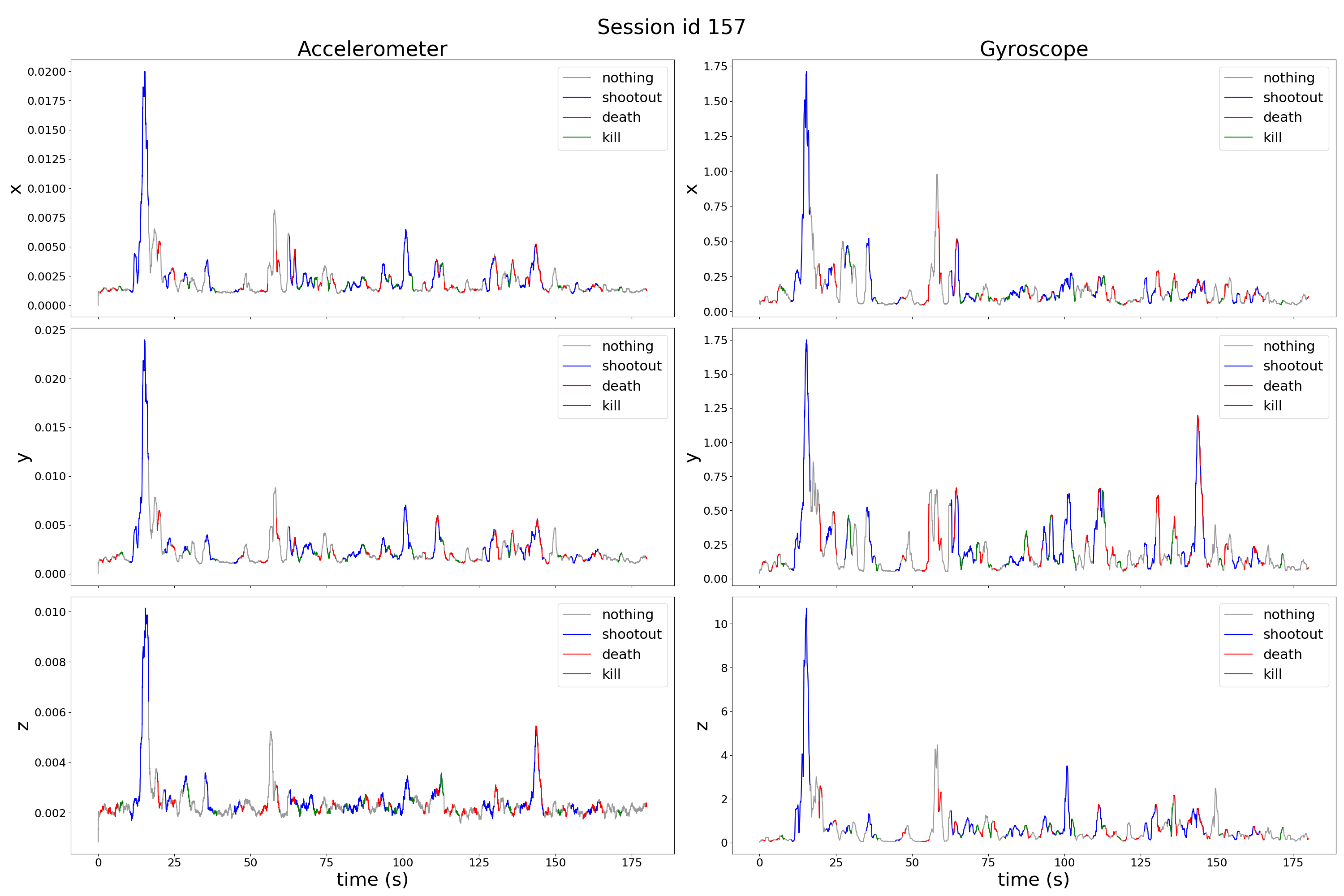}}
	\label{sensors_std_2}
	\end{subfigure}
	\caption{The floating standard deviation of the sensors data within the 1-second window. Colors correspond to the game events.}
	\label{sensors_std}
\end{figure}

While the accelerometer and gyroscope already get  acceleration and angular velocity, their raw measurements do not exactly correspond to the movement due to gravity and non-perfect calibration. These measurements are stationary when the player does not move and nonstationary when the player makes movements. It seems more reasonable, therefore, not to use raw data from the sensors, but to apply increment of the time-series. Thus, in order to more effectively  extract the disturbance, we calculated the standard deviation within the 1-second moving window for each of the sensors. The results obtained are shown in Fig.~\ref{sensors_std}. The peaks correspond to the player's active movement on the chair.

\subsection{Feature Extraction}

We formally define the movement on the chair as  the moments when the floating standard deviation is 3 times larger than its median value. That means that the player probably changes his posture, twitches or leans on the back of the chair.

Then we obtained data on how often persons actively move on the chair as a proportion of these moments. In order to catch the players' reactions to the game events we calculated how often persons actively move within 1 second after killing or death, or during a shootouts.

Another extracted feature is the portion of time when the person leans on the back of the chair. It is easily calculated using that records from the  $z$-component of the accelerometer decreases when the person leans on the back.

In order to assess the player's actual  performance during the session, we obtained the Kill Death Ratio (KDR) from the game logs, a popular metric for estimating the player's skill. If the player did not  die within the session, we bounded this value by 10.

\begin{figure*}[!bt]
	\centerline{\includegraphics[width=\linewidth]{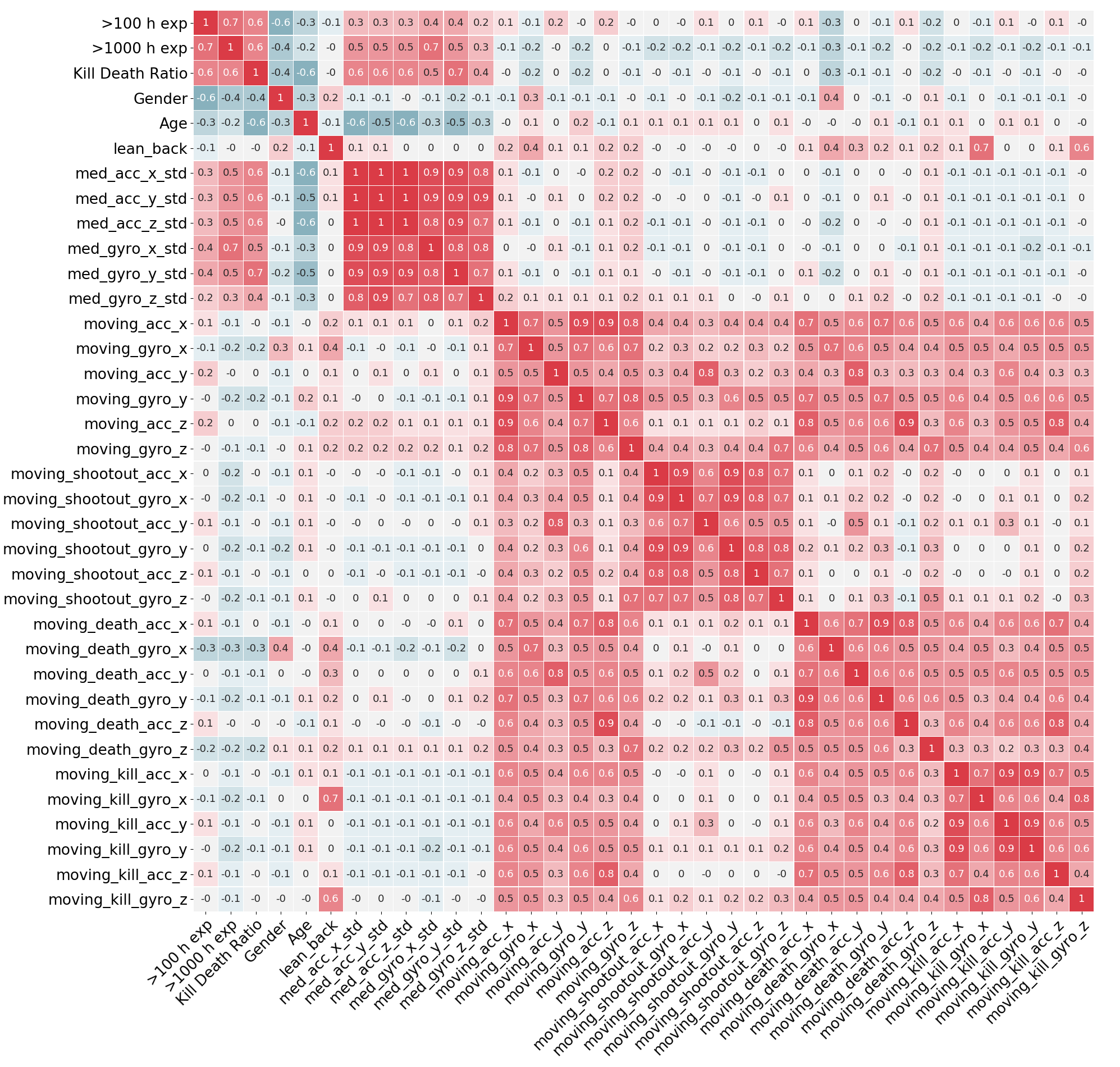}}
	\caption{Correlations between thefeatures.}
	\label{heatmap_features}
\end{figure*}

The correlations between the obtained features are shown in Fig.~\ref{heatmap_features}. Though we added gender and age of the player for illustrative purposes, we did not use these data in further experiments. The features are described in more detail  in Table~\ref{factors_description}.

\begin{table}[!bt]
	\caption{Description of features.}
	\begin{center}
		\begin{tabular}{|p{2.3cm}|p{5.7cm}|}
			\hline
			\textbf{Feature name} & \textbf{Description} \\\hline
			$>$1000 h exp & Player has more than 1000 hours of experience. \\\hline
			Gender & Gender of a player. 0 is a woman, 1 is a man. \\\hline
			Kill Death Ratio & Number of player's kills divided by number of player's death in a session \\\hline
			Age & Age of a player. \\\hline
			lean\_back & Portion of time when player leans to the back of the chair. \\\hline
			med\_acc\_x\_std & Median value of the floating standard deviation within 1-second window for $x$-component of accelerometer. The same for $y$ and $z$ components and gyro.  \\\hline
			moving\_acc\_x & Proportion of time when floating standard deviation of $x$ component of accelerometer is 3 times more than median. In other words, player actively moves along $x$ axis. \\\hline
			moving\_gyro\_x & The same as moving\_acc\_x, but player actively rotates along $x$-axis. \\\hline 
			moving\_death\_acc\_x & Proportion of time during 1 second after death when person actively moving. The same for gyro, other components and events.\\\hline
		\end{tabular}
		\label{factors_description}
	\end{center}
\end{table}

It is clear from the correlation plot that many features obtained from the chair sensors are highly correlated. It is usually the case for the groups of 6 features and can be explained as follows: moving along one direction oftentimes implies moving along other directions.
The heatmap also shows some interesting conclusions  from the data the older players, for example, move on a chair less than the younger players, or that men lean to the back of the chair more often than women.

The reason why we used the fact that the player has more than 1000 hours of experience as a target is that the professional high-skilled player definetely has more than 1000 hours of the game time, while the low-skilled player has probably less than 1000 hours of experience. On top of that this target is more informative than that of 100 hours according to the heatmap. As for KDR which is also a good target, it significantly fluctuates within various experiments:  people on the server are constantly changing and it would be difficult to build a stable machine learning model for this target. Moreover, KDR, being a continuous metric, has to be somehow splitted into several classes to apply classification. However, a particular variant of such splitting is not easy to justify.

\subsection{Feature Selection}

In order to figure out the most important factors and to build a stable machine learning model, the 
feature selection algorithms should be applied.
Assuming that each feature monotonically affects the target and considering high correlations between them,  the feature selection algorithms based on the linear models can be successfully applied. In particular, LASSO method can select the most important factors in a linear model (here we temporary switch over to the  regression problem instead of the classification). However, a number of the features selected strongly depend on a regularization constant, which is selected manually \cite{tibshirani1996regression}. 

LASSO is the method which minimizes the functional:

\begin{equation}
\frac1{2n}\|y - Xw\|_2^2 + \alpha \|w\|_1,
\label{lasso}
\end{equation}
where $X$ is the  design matrix, $y$ is the vector with target values, $w$ is the vector of coefficients, $\alpha$ is the regularization constant.

In order to select the optimal regularization constant $\alpha$ and the most important features, the information criteria, e.g. the Akaike information criterion (AIC) or Bayes Information criterion (BIC),  are widely used \cite{kuha2004aic}. These criteria can efficiently select the accurate model which uses some of the most important features.
It is achieved by simultaneously penalizing the model for a number of used features and maximizing the likelihood function of the model.

The AIC value for the model is determined by the number of the $k$ parameters (the number of features in our case) and the maximum value of the likelihood function $\hat{L}$ with respect to those parameters:

\begin{equation}
\text{AIC} = 2k - 2\ln(\hat{L}).
\end{equation}

The BIC is similar, but it penalizes the number of features more significantly:

\begin{equation}
\text{BIC} = \ln(n)k - 2\ln(\hat{L}),
\end{equation}
where $n$ is the number of observations. The smaller  both AIC and BIC values are, the better.

\begin{figure}[!bt]
	\centerline{\includegraphics[width=\linewidth]{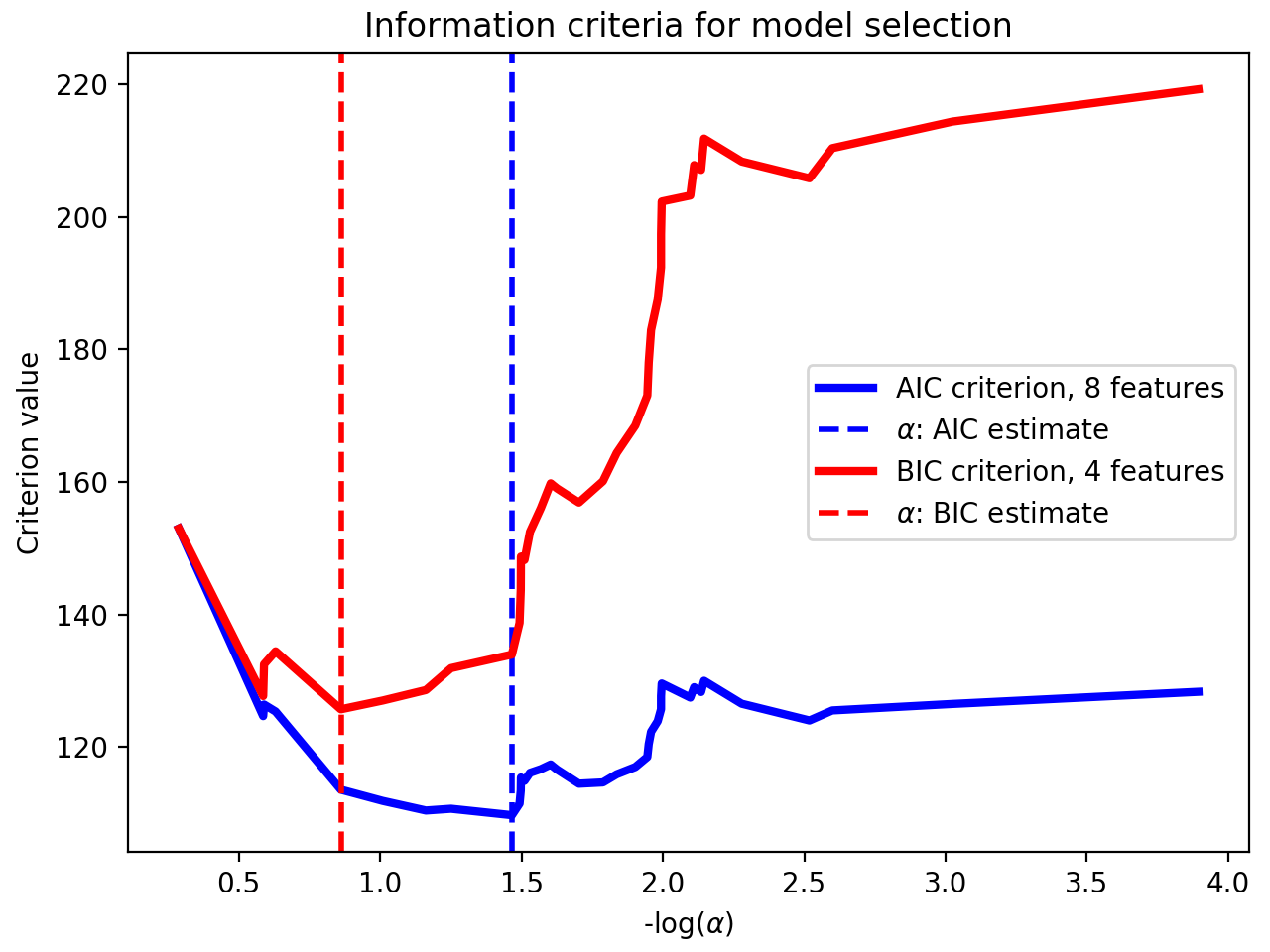}}
	\caption{Dependence of AIC and BIC for LASSO w.r.t. $\alpha$.}
	\label{aic_bic}
\end{figure}

The dependence of AIC and BIC w.r.t. $\alpha$ is shown in Fig.~\ref{aic_bic}. The best model according to AIC has 8 features, while according to BIC it has 4 features - and all of them are included in the top-8 features. These key factors are presented in Table~\ref{aic_features}. The negative coefficients correspond to the factors that are intrinsic to the  low-skilled players, while the positive coefficients correlate with the features that are specific to the  high-skilled players. Coefficients with the higher absolute values are more important.

Interestingly enough, there are no features associated with the event when the player makes a frag. In this situation, apparently all the players react in the same way. Besides, there are no features related to the $z$-component of the accelerometer, perhaps for the reason that the height of the chair is fixed at the beginning of the experiment and does not change further.

\begin{table}[!bt]
	\caption{Features selected using AIC.}
	\begin{center}
		\begin{tabular}{|p{2.9cm}|p{0.7cm}|p{4cm}|}
			\hline
			\textbf{Feature name} & \textbf{Coef.} & \textbf{Possible sense}\\\hline
			moving\_death\_gyro\_x & -0.17 & How often after the death player quickly leans back (or opposite, get close to a monitor).\\\hline
			moving\_shootout\_gyro\_z & -0.16 & How often during the shootout player spins on the chair.\\\hline
			moving\_death\_gyro\_y & -0.07 & After the death player changes a posture.\\\hline
			moving\_shootout\_acc\_y & -0.07 & How often during the shootout player approaches or move away from monitor.\\\hline
			med\_gyro\_y\_std & 0.03 & How intensely player moves in a chair.\\\hline
			moving\_acc\_x & 0.05 & How often player moves along the table.\\\hline
			moving\_acc\_y & 0.11 & How often player approaches or move away from a monitor.\\\hline
			med\_gyro\_x\_std & 0.42 & How intensely player wiggles to a monitor.\\\hline
		\end{tabular}
		\label{aic_features}
	\end{center}
\end{table}

\subsection{Machine Leaning}

In order to estimate the possibility of predicting the  player's skill by his/her behaviour on a chair, we built several machine learning models. To obtain correct results models were trained on all sessions of a random half of players and validated on all sessions of other players. for training we used only 8 features selected above. As the target variable $y$  we use the fact that a person has more than 1000 hours experience in CS:GO.

We applied 6 fundamentally different standard machine learning algorithms with hyperparameters adjusted to our problem:

\begin{enumerate}
	\item Logistic regression. Simple linear model for classification \cite{kleinbaum2002logistic},
	\item Support vector machine (SVM) with radial basis functions (RBF) kernel. Nonlinear method for data separation \cite{amari1999improving},
	\item Random forest with 100 estimators and maximum tree depth 2. Plenty of diverse decision trees voting for the optimal class \cite{liaw2002classification},
	\item k-nearest neighbors classifier with k=3. Simple algorithm trying to find similar objects in the train data \cite{cover1967nearest},
	\item Naive Bayes. This method assumes that the  features are independent and estimates their distribution for each target class. Since the data are mostly continuous, we used Gaussian distribution (Gaussian Naive Bayes) \cite{naive_bayes},
	\item Gaussian process. This algorithm implies the  probabilistic nature of data and tries to predict  the target class using a latent function \cite{gaussian_process}.
\end{enumerate}

To describe methods performance comprehensively we used several evaluation metrics. 

\begin{enumerate}
	\item Accuracy, or proportion of right predictions. Simple metric estimating chance of the right prediction. The higher values are, the better.
	\item The area under the receiver operating characteristic curve (ROC AUC). It ranges from 0 to 1 with the 0.5 value for random guessing. The  higher values are, the better  \cite{fan2006understanding}.
	\item Log Loss, or cross-entropy for binary case. This describes the imperfection of the  classification from the information theory view. The lower values are, the better.
\end{enumerate}

Since the number of positive and negative samples is  approximately the same, the application of all these metrics is reasonable. For the more precise estimation, each metric was calculated for 1000 different random train/test splits and averaged.
The evaluation results for all the algorithms are shown in Table~\ref{scores}.

\begin{table}[!bt]
	\caption{Scores for the machine learning algorithms.}
	\begin{center}
		\begin{tabular}{|c|c|c|c|c|}
			\hline
                       & Accuracy & ROC AUC  & Log Loss \\\hline
Logistic regression     & 0.71      & 0.86 & 0.60    \\\hline
SVM                    & \textbf{0.78}      & 0.85  & 0.84    \\\hline
Random forest & 0.77     & \textbf{0.88} & \textbf{0.46}    \\\hline
k-nearest neighbors   & 0.77      & 0.73 & 5.59 \\\hline
Naive Bayes   & 0.76      & 0.69 & 5.50 \\\hline
Gaussian process   & 0.71      & 0.86 & 0.60 \\\hline
		\end{tabular}
		\label{scores}
	\end{center}
\end{table}

The random forest turned out to be the best performing algorithm, probably because it can catch non-trivial interactions between the features. SVM has a slightly better accuracy, but much worse Log Loss. 0.88 ROC AUC score is an improvement as compared to 0.86 achieved in \cite{smart_chair_skoltech_1} where information about the game event was not used. It means that we can differentiate the high-skilled player from the  low-skilled player with the 78\% accuracy. More important, however, is that this score can be used for the additional estimation of the player skill in professional CS:GO teams in a short period of time.

\begin{figure}[!bt]
	\centerline{\includegraphics[width=\linewidth]{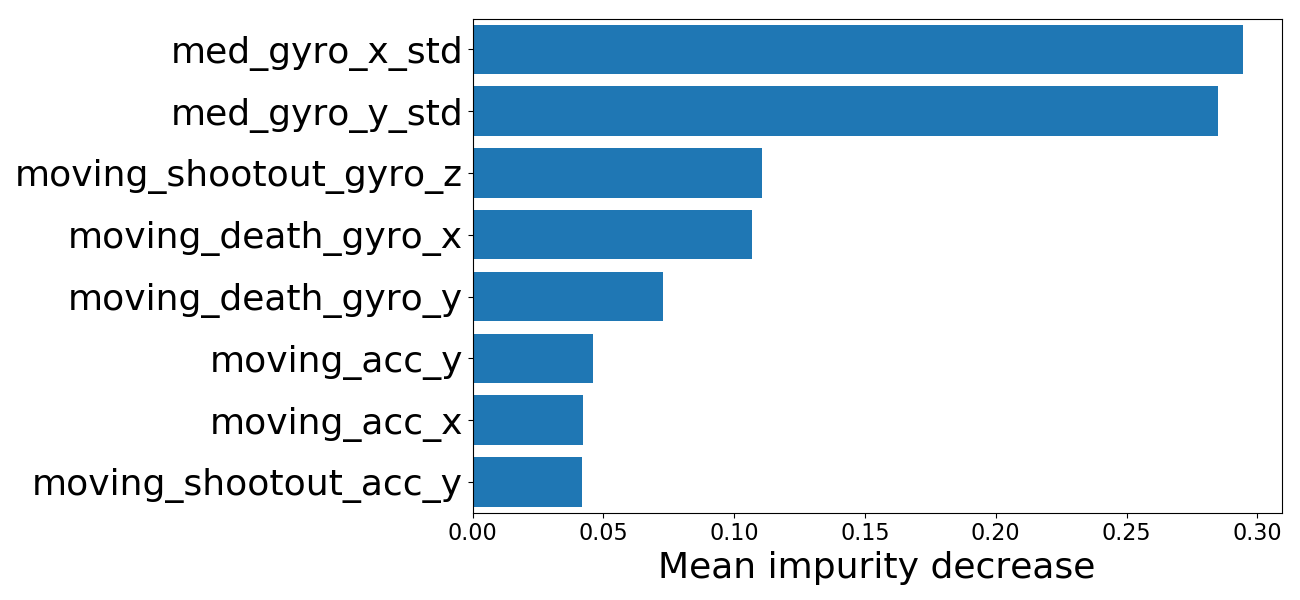}}
	\caption{Feature importance as a mean impurity decrease for random forest.}
	\label{feature_importance_rf}
\end{figure}

In order to determine which of the 8 selected features are more essential, we calculated the feature importance for the random forest based on the mean decrease in impurity. These results are more reliable than estimating the feature importance estimation by combination of AIC and LASSO (see Table~\ref{aic_features}), for the random forest model can catch nonlinear dependencies and performs better according to Table~\ref{scores}.
Results are shown in Fig.~\ref{feature_importance_rf}.

The general activity on the chair turned out to be more important than reactions to game events. Each component from the accelerometer and the gyroscope,  except for the $z$-component of the accelerometer, was useful for determining a player's skill.

\section{Conclusions}

In this work we have proposed the smart chair sensing platform for the collection and analysis of data in eSports. We have processed the raw data from the sensors integrated in the chair and combined them with the key game events.
These data have been further engineered into the features that helped us figure out the most important qualities intrinsic to the professional CS:GO athletes.
Several machine learning models have been built to assess informativeness of the data. As a result, we have got the accurate algorithm capable of predicting the player's skill exclusively on the basis of information received within approximately 3 minutes of his game session. 

Future work provides for the more accurate data preprocessing and timeseries segmentation using hidden markov models \cite{Market2009} and anomaly detection approaches \cite{kNN2017}. The online prediction of the player's performance shall be carried out using the specific metrics for classification of time-series segments \cite{Vehicle2017} and manifold learning for nonlinear feature extraction \cite{DRreg}.

\section*{Acknowledgment}
The reported study was funded by RFBR according to the research project No. 18-29-22077\textbackslash 18.

Authors would like to thank Skoltech Cyberacademy, CS:GO Monolith team and their coach Rustam ``TsaGa'' Tsagolov for fruitful discussions while preparing this article. Also, the authors thank Alexey ``ub1que'' Polivanov for supporting the experiments by providing a slot at the CS:GO Online Retake server (http://ub1que.ru).

\bibliography{main}{}
\bibliographystyle{IEEEtran}

\end{document}